\documentclass[%
preprint,
showpacs,
showkeys,
preprintnumbers,
nofootinbib,
amsmath,amssymb,amsfonts,
aps,
prd,
floatfix,a4paper,11pt]{revtex4-1}
\usepackage{}
\usepackage{cases}
\usepackage{mathrsfs}
\usepackage{amsmath}
\usepackage{amssymb}
\usepackage{color}
\usepackage{CJK}
\usepackage{graphicx}
\usepackage{dcolumn}
\usepackage{bm}
\usepackage[dvipdfm,bookmarks=true,colorlinks,%
            citecolor=blue,linkcolor=blue, hypertex, %
            breaklinks=true]{hyperref}

\begin{document}
\preprint{MPP-2012-81}
\title{1+1-dimensional p-wave superconductors from intersecting D-branes}
\date{\today}
\author{Yanyan Bu}
\email{yybu@mpp.mpg.de}
\affiliation{State Key Laboratory of Theoretical Physics, Institute of
Theoretical Physics, Chinese Academy of Science, Beijing 100190, People's
Republic of China}
\affiliation{Max-Planck-Institut f$\ddot{u}$r Physik
(Werner-Heisenberg-Institut),
F$\ddot{o}$hringer Ring 6, 80805 M$\ddot{u}$nchen, Germany}
\begin{abstract}
In this work we explore $1+1$ dimensional p-wave superconductors using the probe D-brane construction. Specifically, we choose three intersecting D-brane models: D1/D5, D2/D4 and D3/D3 systems. According to the dilaton running behavior, we denote the former two systems as nonconformal models and the last system as conformal. We find that all three models are qualitatively similar in describing superconducting condensate as well as some basic features (such as the gap formation and DC superconductivity) of superconducting conductivity. There also exist some differences among the three models as far as the AC conductivity is concerned. Specifically, for D3/D3 model there is no peak at nonzero frequency for the imaginary part of the conductivity, which is present in the nonconformal models; their asymptotic behaviors are different---for D1/D5 the real part of the AC conductivity approaches one at large frequency limit, for D2/D4 it slowly goes to a certain nonzero constant smaller than one and for D3/D3 it goes to zero. We find that the profile of the AC conductivity for the D1/D5 system is very similar to that of higher dimensional p-wave superconductors.
\end{abstract}
\pacs{11.25.Tq}
\maketitle
\tableofcontents
\section{Introduction} \label{section1}
The applications of AdS/CFT correspondence \cite{hep-th/9711200}, more generally gauge/gravity duality, to investigations of strongly coupled system have gained broad interest varying from QCD phenomena at low energy to strongly correlated
condensed matter physics, see, e.g., \cite{0711.4467,*0901.0935,*0903.3246,*0904.1975,*0904.2750,*0909.0518} for recent reviews. One of the most interesting applications is the construction of superconducting-like phase transition. Following the pioneering work of \cite{0801.2977,*0803.3483}, holographic superconductors have been constructed
in \cite{0803.3295,*0810.1563,0805.2960,0805.3898} by putting the Abelian Higgs model or SU(2) gauge field into the AdS black hole geometry. When the Hawking temperature is decreased to some critical value, the black hole background becomes unstable against perturbations and gets hair by condensing some field in order to cure the instability. This can be considered as holographic realization of the superconducting phase transition. This kind of construction of holographic superconductors takes the (asymptotically) AdS black hole spacetime as the starting point. In some sense, this construction should be taken as a bottom-up approach to the gravity dual of strongly interacting superconductor as the theory in the bulk is directly written down from phenomenological point of view. Recently, there appear some works on the UV completion of these phenomenological superconducting models, see, e.g.,
\cite{0901.1160,*0907.3510,*0907.3796,*0912.0512,*1011.2172,*1104.4478,*1110.3454}, by
embedding the holographic superconductors into the superstring/M-theory or gauged supergravity. Another top-down approach to holographic superconductors where the dual field theory is known is based on probe D-brane in a black p-brane supergravity geometry. Such a holographic superconductor has been established in \cite{0810.2316,*0903.1864,*0810.3970} where there is a $\rho$ meson condensate. Moreover, a stringy mechanism for the condensation process has been described. For the Sakai-Sugimoto model \cite{hep-th/0412141}, a holographic superconductor involving a $\rho$ meson condensate has been described in \cite{0907.1508}, based on earlier results on $\rho$ meson condensation in \cite{0709.3948}. In \cite{1011.5218} the D3/D5 system,
constructed to be a defect theory, was used to study different condensates corresponding to vector/scalar modes on the flavor D5 brane worldvolume.

The models mentioned above are all concerned on higher dimensional spacetime, say planar or 1+3-dimensional superconductors. However, 1+1 dimensional system is also of great importance and interest in condensed matter physics. It is therefore of large interest to see what holographic methods tell us about such systems. The work of \cite {0909.3526} takes the D3/D3 brane system (This model was first studied in \cite{hep-th/0211222, *hep-th/0512125}) to model 1+1 dimensional strongly coupled quantum liquid and found some interesting properties different from higher dimensional counterparts. Similar studies
taking the BTZ black hole geometry as the gravity background can be found in \cite{0909.4051}. Some aspects of holographic Luttinger theorem have been discussed in detail in \cite{1203.5388}. More recently, holographic s-wave superconductors in 1+1 dimensions have been constructed and well-studied in
\cite{1008.3904,1011.3520,1106.4353,*1107.2909} by introducing Maxwell-Scalar system into the BTZ black hole geometry. Other recent works for 1+1 dimensional boundary theory from holography can be found in \cite{1008.4350,1012.4831,1103.6286}. In \cite{1204.3103} local aspects of 1+1-dimensional superconductivity, which followed the pioneering work of \cite{1005.1776,*1202.0006}, have been investigated in a bottom-up
approach. This ensures a true superconductor where a local symmetry is spontaneously broken instead of a superfluid where a global symmetry is broken.

In this paper we take the probe D-brane approach to study some aspects of p-wave superconductor in 1+1 dimensional spacetime. More specifically, we take three intersecting D-brane systems as our starting point: D1/D5, D2/D4 and D3/D3. For stability of these brane systems, we keep them supersymmetric in the sense that the ND number of each system is 4. For simplification, our study is limited to the zero quark mass and probe limit (i.e., the backreaction of the flavor probe brane on the background geometry is neglect). As in \cite{0810.2316,*0903.1864,*0810.3970}, we embed two coincident probe D-branes into the black p-brane geometry and just reserve the Yang-Mills truncation of
nonlinear DBI action for probe D-brane. As mentioned earlier, one advantage of taking the intersecting D-brane system is that its field theoretical side is known, here they are supersymmetric gauge theories coupled with fundamental matters. The superconducting condensates are nonabelian gauge fields, living on the probe flavor branes, and have one Lorentzian index, indicating the phase transition is of the p-wave type. Our numerical results for the D3/D3 model are very similar to those of \cite{1204.3103}. However, one main advantage of the D3/D3 system is that its dual field theory is known (the defect 1+1 dimensional CFT) and the condensed operator can be explicitly written down, see eq.~(\ref{condensate operator}).

Although these probe D-brane constructions for p-wave superconductor seem very general as well as similar for different D-branes, the results are not always the same. Specifically, we find that nonabelian condensates for the three models are qualitatively the same, they have mean field behavior near the critical temperature and approach to some fixed constants at very low temperature. When turning to the electromagnetic response, we find that, for the real parts of the conductivities, the three models give some features in common, like the DC infinite conductivity and the gap formation. However, for D3/D3 model, there is no peak at nonzero frequency for imaginary part of the AC conductivity compared to the others. Actually, the conductivity formula for the D3/D3 system is very different from the other two models due to the nontrivial behavior for the electromagnetic fluctuations near the AdS boundary. Another main difference among these models is that their asymptotic behaviors of the AC conductivity are quite different. The real part of the AC conductivity approaches one at large frequency limit for D1/D5 system and goes to a nonzero constant (much smaller than one and also very slowly) for D2/D4 model while for D3/D3 system it tends to zero.

The rest of this work is organized as follows. In section~\ref{section2}, we shortly review previous studies on quantum field theory in 1+1 dimensional spacetime by taking the approach. We then introduce the models studied in this work and numerically solve the nonlinear equations of motion for background fields. With these numerical solutions, we plot the superconducting condensate as well as the free energy versus the dimensionless temperature. Section~\ref{section3} is devoted to the study of the electromagnetic fluctuation. We plot the AC conductivities for all the three systems and give some comments on the results. We end with a short summary and some discussions in section~\ref{section4}.
\section{Flavor p-wave superconductors from intersecting D-branes}\label{section2}
\subsection{Some reviews}
In this subsection, we shortly review previous studies about 1+1 dimensional quantum field theory by taking the AdS/CFT approach. The first aspect we intend to mention here is about holographic quantum liquids in 1+1 dimensions, initially studied in \cite{0909.3526,0909.4051}. Although the authors in \cite{0909.3526} take defect D3/D3 intersecting D-brane models as the starting point, their results are in quite agreement with those of \cite{0909.4051}, which directly takes the charged BTZ black hole to compute correlation functions for probe scalar, spinor and vector operators in this geometry. In contrast with higher dimensional defects, a persistent dissipationless zero sound mode is found in \cite{0909.3526}. The correlation between log periodicity and the presence of finite spectral density of gapless modes is seen in \cite{0909.4051}. Meanwhile, the real part of the conductivity (given by the current-current correlator) also vanishes as $\omega \rightarrow 0$ as expected. The fermionic Green's function shows quasiparticle peaks with approximately linear dispersion but the detailed structure is neither Fermi liquid nor Luttinger liquid and bears some similarity to a "Fermi-Luttinger" liquid. As will be seen in later sections, to some degree, our numerical results for the AC conductivity in D3/D3 model is in agreement with these conclusions.

The second point is about holographic realization of symmetry breaking in 1+1 dimensional spacetime, which has been discussed in \cite{1008.3904,1011.3520}. The authors in \cite{1011.3520} considered a system of 3D gravity coupled to matter to study the symmetry breaking phases in 1+1 dimensional spacetime. To be specific, they model symmetry breaking phases of a strongly coupled 1+1 dimensional CFT as black holes with scalar hair. It concluded that, in the case of a discrete symmetry, these theories admit metastable phases of broken symmetry. Moreover, the 3D Einstein-Maxwell theory shows continuous symmetry breaking at low temperature. The latter conclusion can be used to construct holographic s-wave superconductors in 1+1 dimensional spacetime. Intuitively, continuous symmetry breaking in 1+1 dimensional spacetime seems to contradict with the
Coleman-Mermin-Wagner (CMW) theorem
\cite{10.1103/PhysRevLett.17.1133,*10.1103/PhysRev.158.383,*10.1007/BF01646487}
which states that in 1+1 and 2+1 space-time dimensions at finite temperature, spontaneous continuous symmetry breaking is impossible. However, in the large N limit, these lower dimensional systems can have another phase in which the continuous symmetry is almost spontaneously broken and the fall of the correlation functions is of the power law type as pointed out in \cite{10.1016/0550-3213(78)90416-9}. Under the framework of the AdS/CFT correspondence, one may expect that $\mathcal{O}(1/N)$ corrections will wash out the symmetry breaking phase. Actually, this indeed happens as was explored in \cite{1005.1973} for the $AdS_4$ black hole. It is then reasonable to believe that this idea also holds in the three dimensional bulk case. Given these facts, it is very interesting to see what holographic method can tell us about the superconducting phase transition in two dimensional spacetime with vector order parameter. What we found is
that, when working in the large N limit, which is the fundamental assumption of
AdS/CFT correspondence, holographic superconductor symmetry breaking in 1+1-dimensional spacetime does happen.

Another interesting feature for 3-dimensional bulk theory is the chiral anomaly, which has been studied in \cite{1012.4831} with applications to condensed matter physics and in \cite{1103.6286} for correctly producing meson spectrum. In the bulk side, the chiral anomaly is the Chern-Simons term in the bulk action. The author in \cite{1012.4831} studied the holographic description of finite-density systems in two dimensions. Quite interesting, it was shown that the chiral anomaly for symmetry currents in 2 dimensional CFT completely determines their correlators. The important exception is a CFT with a gauge theory to which we may couple an external current, as in the probe D3/D3 system or the putative dual to the charged BTZ black hole. In \cite{1103.6286}, the defect D2/D8 brane model was used to holographically realize large Nc massless QCD in two dimensional spacetime. The flavor axial anomaly is dual to a three dimensional Chern-Simons term which turns out to be of leading order, and it affects the meson spectrum and holographic renormalization in crucial ways. It was also shown that an external dynamical photon acquires a mass through the three dimensional Chern-Simons term as expected from the Schwinger mechanism. Massless two dimensional QCD at large Nc exhibits anti-vector-meson dominance due to the axial anomaly. Explicitly, one still cannot give a general argument on the effect of the chiral anomaly term on the dynamics. However, the models studied in this work do not contain the chiral anomaly terms, which greatly simplify our analysis as well as numerical computations. If we go beyond the massless limit for the flavor quark, the Chern-Simons (corresponding to chiral anomaly) will appear and we leave the study along this line for future work.

Before concluding this subsection, we briefly comment the models used in this work. Although the three models considered here look quite similar, there are still some differences among them, which result in different features for the AC conductivity, which we will reveal in later sections. The D1/D5 and D2/D4 models are nonconformal in the sense that the dilaton profiles are non-trivial. Then, according to the AdS/CFT correspondence, the gauge coupling constant on the dual field theory side run along the radial direction, which can be thought of as the energy scale. We may naively denote these two models as the nonconformal SYM gauge theory coupled to fundamental matter. However, the D3/D3 model is a conformal one. These differences together with the specific background spacetime tell us that the gauge field living on the probe D-brane has quite different asymptotic behavior (here we mean the behavior at the AdS conformal boundary) for these models. In particular, there is a logarithmic term for the boundary expansion of gauge field in the D3/D3 case, which will be clear in \ref{subsection1}. This makes one to identify the coefficient of the logarithmic term instead of the constant term as the source. This never happens for higher dimensional (asymptotic) AdS geometry case\footnote{However, this happens in choosing the Lifshitz black hole geometry to construct the strong coupling superconductor with dynamical exponent, as seen in \cite{10.1103/PhysRevD.86.046007}.}. Furthermore, due to this point, one have to give a new prescription for the computation of the transport coefficients in D3/D3 system. The details on these issues will be given in later sections. In \cite{0810.2316,*0903.1864,*0810.3970,0907.1508}, the authors mainly focused on the decoupled sector\footnote{The coupled sector in the probe D-brane setup has been revealed recently in \cite{10.1016/j.nuclphysb.2012.07.014}.}, which has similar equation of motion as in the s-wave case. However, here we directly encounter the coupled modes, and need to define a gauge invariant variable to plot the conductivity. This makes our numerical computations complex, especially for the D3/D3 system.
\subsection{Equation of motion for the background fields}
The gravity dual of p-wave superconductor was first constructed in \cite{0805.2960,0805.3898} by putting the SU(2) gauge field into the $AdS_4$ black hole geometry. The action for this gravity system is
\begin{equation}
S=\frac{1}{2\kappa^2}\int d^4x\sqrt{-g}[R-2\Lambda-\frac{1}{4}F_{\mu\nu}^aF^{a\mu\nu}],
\end{equation}
where the field strength tensor of nonabelian SU(2) gauge field $A_{\mu}^a$ is defined as $F_{\mu\nu}^a=\partial_{\mu}A_{\nu}^a-\partial_{\nu}A_{\mu}^a+\epsilon^{abc}A_{\mu}^b A_{\nu}^c$ with totally antisymmetric tensor $\epsilon^{123}=+1$. Working in the probe limit allows us to ignore the backreaction of the gauge field on the background geometry. Therefore, we can fix the black hole geometry as the Schwarzschild-AdS spacetime and study nonabelian gauge field in this curved geometry. The anisotropic features of the p-wave superconductor were explicitly reflected on different behaviors of the conductivities along $x$ and $y$ directions. Followed by this work are some developments of holographic p-wave superconductors, see
\cite{0902.0409,*0906.2323,*0911.4999,*0912.3515,*0912.4928,*1002.4416,*1003.1134,
*1007.3321,*1011.5912,*1012.5559,*1103.3232,*1106.0784,*1109.4592,*1110.0007} for an incomplete list.

However, all of these works are taking higher dimensional AdS black hole geometry as the starting point. We here use the probe brane method, which is first applied to holographic p-wave superconductor in \cite{0810.2316,*0903.1864,*0810.3970}, to explore some properties of 1+1 dimensional p-wave superconductors. We choose the p-brane geometry as our background metric,
\begin{equation}
ds_p^2=H^{-1/2}\left(-f(\rho)dt^2+d\vec{x}^2\right)+H^{1/2}\left(\frac{d\rho^2}{
f(\rho)}+\rho^2 d\Omega_{8-p}^2\right),
\end{equation}
with
\begin{equation}
e^{\phi}=H^{\frac{3-p}{4}},\quad H=\left(\frac{L}{\rho}\right)^{7-p},\quad
f(\rho)=1-\left(\frac{\rho_0}{\rho}\right)^{7-p},
\end{equation}
where $\vec{x}=\left(x^1, x^2,\cdots,x^p\right)$ denotes the p-dimensional space of the black p-brane. Explicitly there is a horizon in above metric at $\rho=\rho_0$ and the Hawking temperature is
\begin{equation}
T=\frac{7-p}{4\pi L}\left(\frac{\rho_0}{L}\right)^{\frac{5-p}{2}}.
\end{equation}
For our purpose, we will choose $p=1,2,3$ and parameterize the internal spaces $\Omega_{8-p}$ as follows,
\begin{eqnarray}
&d\Omega_7^2=d\theta^2+\cos^2\theta dS_3^2+\sin ^2\theta d{S_3^{\prime}}^2,\nonumber\\
&d\Omega_6^2=d\theta^2+\cos^2\theta dS_2^2+\sin ^2\theta dS_3^2,\nonumber\\
&d\Omega_5^2=d\theta^2+\cos^2\theta d\xi^2+\sin ^2\theta dS_3^2.
\end{eqnarray}
The probe D-brane extends along $\left(t, x^1, \rho, S_3\right)$ for D1/D5 model, $\left(t, x^1, \rho, S_2\right)$ for D2/D4 model and $\left(t, x^1, \rho, \xi\right)$ for D3/D3 system, respectively. The embedding profile for the probe D-brane can be parameterized by $\theta(\rho)$. As mentioned before, we here consider the zero quark mass case, i.e., $\theta(\rho)=0$ and leave the effect of the nonzero quark mass for future research. With these assumptions, the induced metrics on the flavor probes are
\begin{eqnarray}
&ds_5^2=\left(\frac{\rho_0}{L}\right)^3\frac{1}{u^3} \left(-f(u)dt^2+dx^2\right)+\frac{L^2}
{\rho_0u}\frac{du^2}{f(u)}+\frac{L^3}{\rho_0}u dS_3^2,\quad
f(u)=1-u^6,\nonumber\\
&ds_4^2=\left(\frac{\rho_0}{L}\right)^{5/2}u^{-5/2}\left(-f(u)dt^2+dx^2\right)+
\frac{L^{5/2}}{\rho_0^{1/2}}u^{-3/2}\frac{du^2}{f(u)}+\frac{L^{5/2}}{\rho_0^{1/2
}}u^{1/2} dS_2^2,\quad f(u)=1-u^5,\nonumber\\
&ds_3^2=\left(\frac{\rho_0}{L}\right)^2\frac{1}{u^2}
\left(-f(u)dt^2+dx^2\right)+\frac{L^2}{u^2}
\frac{du^2}{f(u)}+L^2 d\xi^2,\quad f(u)=1-u^4,\label{induced metric}
\end{eqnarray}
where in above formulae we have transformed holographic coordinate $\rho$ to a finite interval [0,1] by transformation $\rho_0/\rho=u$ because we found it is more convenient to work with $u$ coordinate for numerical computations. In this new coordinate system, $u=0$ denotes the AdS conformal boundary where the dual field theory lives and the horizon is located at $u=1$.

Two coincident probe D-branes in above black hole geometries have U(2) gauge symmetry on its worldvolume. However, we in this work concentrate on its nonabelian subgroup SU(2) for the purpose of inducing a p-wave superconducting phase transition. In the probe limit, dynamics of the probe Dq-brane is fully determined by nonabelian DBI action\footnote{Note that, for the intersecting D-brane models considered here, there is no Chern-Simons term contribution to the D-brane action. For one thing, we take the zero mass limit for the probe D-brane embedding profile; for another, we do not consider the excitation of the gauge field along the internal space.} and we only reserve the Yang-Mills truncation of it,
\begin{equation}
S=-\frac{T_q N_f}{4}\int d^{q+1}x e^{-\phi}\sqrt{-g}F_{\mu\nu}^aF^{a\mu\nu},
\label{action}
\end{equation}
with the determinant $g$ calculated from the induced metric in eq.~(\ref{induced metric}). One further assumption which will simplify our computation is that we do not consider the internal coordinates dependence of the SU(2) gauge field. Then we can integrate out the internal space in eq.~(\ref{action})
\begin{equation}
S=-\mathcal{N}_q\int d^3x\sqrt{-G}F_{\mu\nu}^aF^{a\mu\nu},\label{actionnew}
\end{equation}
where $\sqrt{-G}=\sqrt{-g}e^{-\phi}g_S$ with $g_S$ the internal metric and $\mathcal{N}_q$ is a model dependent factor, which is irrelevant for later computations. The equation of motion from this action is of the form,
\begin{equation}
\partial_{\mu}[\sqrt{-G}F^{a\mu\nu}]+\sqrt{-G}\epsilon^{abc}A_{\mu}^b
F^{c\mu\nu}=0.\label{eq motion}
\end{equation}

We consider the chemical potential induced superconducting like phase transition. To achieve this goal, we turn on one time component of nonabelian SU(2) gauge field, say $A_0^3(u)\neq 0$ (here, we only consider homogenous holographic superconductor, i.e , the background $A_0^3$ and $A_1^1$ only depend on holographic coordinate $u$.). Similar to the arguments in \cite{0801.2977,*0803.3483}, one can show that when the chemical potential (whose meaning will be clear later), provided by the nonnormalizable mode of
$A_0^3$, is increased to some critical value the black hole will get unstable against perturbations. This instability can be cured by condensing some component of the gauge field $A_{\mu}^a$, say $A_1^1(u)$. More specifically, we consider following ansatz\footnote{Another configuration for the background has been considered in \cite{0805.2960,0805.3898}. But it is unstable and have a higher energy compared to the one in eq.~(\ref{condensate}).}for the hairy black hole,
\begin{equation}
A=A_0^3\tau^3dt+A_1^1\tau^1dx.\label{condensate}
\end{equation}
As emphasized in the first two references of \cite{0810.2316,*0903.1864,*0810.3970}, the solution with nontrival profile for $A_1^1$ will be the new ground state when $T<T_c$ and this new ground state can be interpreted as a $\rho$ meson superfluid. In what follows we take the D3/D3 system as an example to explicitly write down the condensate operator from field theory point of view. More detailed discussions can be found in the first two refs of \cite{0810.2316,*0903.1864,*0810.3970}. The isospin chemical potential, provided by nonnormalizable mode of $A_0^3$, is introduced as source of the operator
\begin{equation}
J_0^3 \propto \bar{\psi} \sigma^3\gamma_0\psi+\phi\sigma^3\partial_0\phi
\end{equation}
where $\psi=(\psi_u,\psi_d)$ and $\phi=(\phi_u,\phi_d)$ are fundamental quarks and squarks under gauge group SU(2); $\sigma^i$ denotes the Pauli matrices and $\gamma_{\mu}$ the Dirac matrices in two dimensional spacetime. In the same way, the condensate operator, holographically provided by normalizable mode of bulk field $A_1^1$, takes the following form
\begin{equation}
J_1^1 \propto
\bar{\psi}\sigma^1\gamma_1\psi+\phi\sigma^1\partial_1\phi.\label{condensate
operator}
\end{equation}

The condensate in eq.(\ref{condensate}) breaks both the SU(2) and translational invariance in the bulk completely. However, one should keep in mind that the broken symmetry considered here is the flvaor symmetry, like that of the QCD theory, which is somewhat different from the symmetry being broken in studies such as \cite{0805.2960,0805.3898}. The symmetry breaking pattern going through the superconducting phase transition can be understood in the following way. The nonzero value of $A_0^3$ at the AdS conformal boundary explicitly breaks the SU(2) to its subgroup $U(1)_3$, generated by rotation in the colored 12-plane. We can identify this residual unbroken symmetry as the electromagnetic symmetry. Strictly speaking, this identification is not right because gauge symmetry in the bulk corresponds to a global symmetry on the boundary field theory side. However, this model can produce many superconductor-like features. We therefore ignore this difference and just go ahead. The U(1) symmetry should be spontaneously broken when going through a superconducting phase transition. It is implemented by the nonzero expectation value for the operator $\mathcal{O}$ dual to $A_1^1$. Additionally, we need impose the source for $\mathcal{O}$ to be zero for spontaneous symmetry breaking.

Plugging the ansatz (\ref{condensate}) into the equation of motion (\ref{eq motion}) results in
\begin{equation} \label{eq15}
\left\{ \begin{aligned}
&\phi^{\prime\prime}-\frac{1}{u}\phi^{\prime}-\frac{u^2}{f(u)}\psi^2\phi=0,\\
&\psi^{\prime\prime}+\left[\frac{f^{\prime}(u)}{f(u)}-\frac{1}{u}\right]\psi^{
\prime}+\frac{u^2}
{f^2(u)}\phi^2\psi=0
\end{aligned} \right.
\end{equation}
where $\left(\phi,\psi\right)\equiv\frac{3}{2\pi T}\left(A_0^3, A_1^1\right)$ for D1/D5 model. Similar results for D2/D4 and D3/D3 systems are listed as below,
\begin{equation} \label{eq24}
\left\{ \begin{aligned}
&\phi^{\prime\prime}-\frac{u}{f(u)}\psi^2\phi=0,\\
&\psi^{\prime\prime}+\frac{f^{\prime}(u)}{f(u)}\psi^{\prime}+\frac{u}{f^2(u)}
\phi^2\psi=0
\end{aligned} \right.
\end{equation}
with $\left(\phi,\psi\right)\equiv\frac{5}{4\pi T}\left(A_0^3, A_1^1\right)$ for D2/D4 and
\begin{equation} \label{eq33}
\left\{ \begin{aligned}
&\phi^{\prime\prime}+\frac{1}{u}\phi^{\prime}-\frac{1}{f(u)}\psi^2\phi=0,\\
&\psi^{\prime\prime}+\left[\frac{f^{\prime}(u)}{f(u)}+\frac{1}{u}\right]\psi^{
\prime}+
\frac{1}{f^2(u)}\phi^2\psi=0
\end{aligned} \right.
\end{equation}
with $\left(\phi,\psi\right)\equiv\frac{1}{\pi T}\left(A_0^3, A_1^1\right)$ for D3/D3 one. In above equations, the prime denotes derivative with respect to $u$ and this notation convention will be used in later representations.
\subsection{Solution for the background fields}\label{subsection1}
Due to the nonlinear coupling between $\phi$ and $\psi$ in eqs.~(\ref{eq15},\ref{eq24},\ref{eq33}), we turn to a numerical shooting method to solve them. The philosophy of our numerical approach is that we first find a power series solution for $\psi$ and $\phi$ near the horizon. We then take these near horizon solutions as initial conditions to numerically integrate these fields from the horizon to the conformal boundary. To the conformal boundary (here, represented by $u=0$), we impose that the source for the operator $\mathcal{O}$ is zero. This condition can filter out the wanted solutions.

We now have a look at the asymptotic behavior of the background fields $\phi$ and $\psi$. Near the conformal boundary, we can get the following asymptotic behaviors from Frobenius analysis of eqs.~(\ref{eq15},\ref{eq24},\ref{eq33}) near the singularity $u=0$,
\begin{eqnarray}
&D1/D5:\quad \phi(u\rightarrow 0)\sim \mu+ \rho u^2+\cdots,\quad
\psi(u\rightarrow 0)\sim \psi^{(0)}+\psi^{(1)}u^2+\cdots,\nonumber\\
&D2/D4:\quad \phi(u\rightarrow 0)\sim \mu+ \rho u +\cdots,\quad
\psi(u\rightarrow 0)\sim \psi^{(0)}+\psi^{(1)}u +\cdots,\nonumber\\
&D3/D3:\quad \phi(u\rightarrow 0)\sim \rho+ \mu \log u +\cdots,\quad
\psi(u\rightarrow 0)\sim \psi^{(1)}+\psi^{(0)}\log u +\cdots.
\end{eqnarray}
Notice that the asymptotic behaviors for $\phi$ and $\psi$ between the conformal and nonconformal models are different---there is a logarithmic term for the D3/D3 model, which also happens in later fluctuation analysis when investigating the electromagnetic response. This difference makes the identification of the source and the operator very different from conventional cases. Here, for the D3/D3 model, the constant term $\psi^{(1)}$ is identified as the operator expectation value because it is now normalizable with respect to the logarithmal term while $\psi^{(0)}$ as the source. Actually, due to this fact, the formula for the AC conductivity of the D3/D3 model is also different from the other two models. As mentioned before, to make the phase transition a spontaneous symmetry breaking, we should impose
\begin{equation}
\psi^{(0)}=0,\quad \psi^{(1)}\sim \langle\mathcal{O}\rangle.
\end{equation}
For the $\phi$ field, the chemical potential $\mu$ should be nonzero and the charge density $\rho$ is a function of it. Actually, D3/D3 model has been exhaustively analyzed in the first reference of \cite{hep-th/0211222, *hep-th/0512125} from the field theory point of view. The action for this system is most easily and elegantly constructed in (2,2) superspace. Different modes on probe D3-brane have also been studied there. However, one main difference from our work is that it also takes into account the $\xi$ (internal space in eq.~(\ref{induced metric})) dependence of flavor U(1) gauge field\footnote{Since we are interested in the lowest energy state, it is reasonable to ignore the $\xi$ dependence for flavor gauge field.}. Therefore, the asymptotic behavior
for gauge modes near conformal boundary is different from our results.

Near the horizon $u=1$, one must have $\phi(1)=0$ for its norm to be finite while $\psi$ should be finite there. We then have following Frobenius expansions for $\phi$ and $\psi$ near the horizon,
\begin{eqnarray}
&\phi(u\sim 1)=a_1(u-1)+b_1(u-1)^2+c_1(u-1)^3+d_1(u-1)^4+\cdots,\nonumber\\
&\psi(u\sim 1)=a_2 + b_2(u-1)+c_2(u-1)^2+d_2(u-1)^3+e_2(u-1)^4+\cdots
\end{eqnarray}
where the coefficients $b_i$, etc. can be uniquely determined in terms of $a_1$ and $a_2$ once plugging these expansion into eqs.~(\ref{eq15},\ref{eq24},\ref{eq33}) and counting by order of $(u-1)$. In carrying out numerical computations, one first guesses some values for $a_1$ and $a_2$. Then one uses this near horizon expansion to apply a finite-element differential equation solving method. At the boundary $u=0$, the condition $\psi^{(0)}=0$ will filter out the wanted solutions. As in \cite{0805.2960}, we will restrict to the solutions where $\psi$ has no nodes because on general grounds the solutions with nodes are expected to be thermodynamically unstable and have higher energy.

\begin{figure}[h]
\includegraphics[scale=0.8]{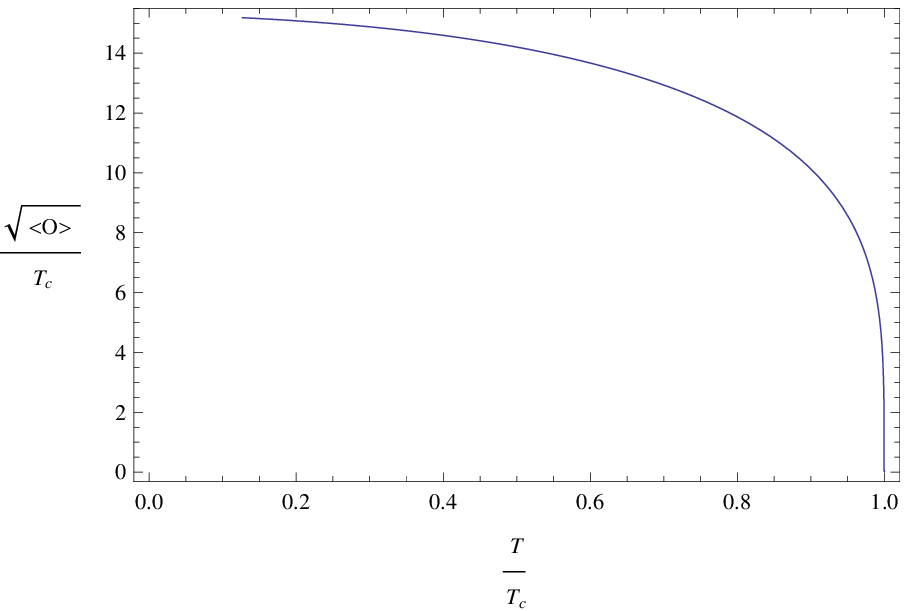}
\includegraphics[scale=0.77]{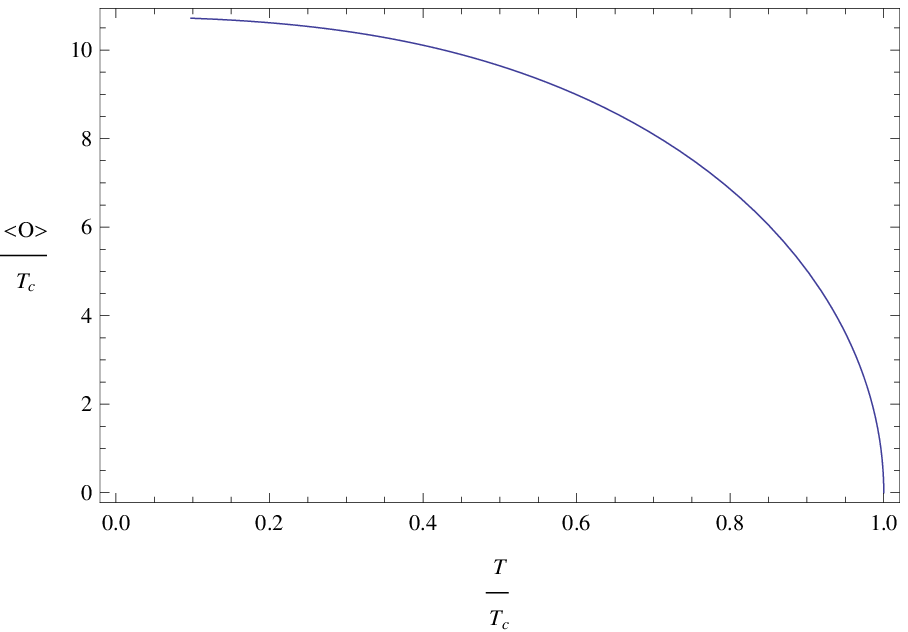}
\includegraphics[scale=0.8]{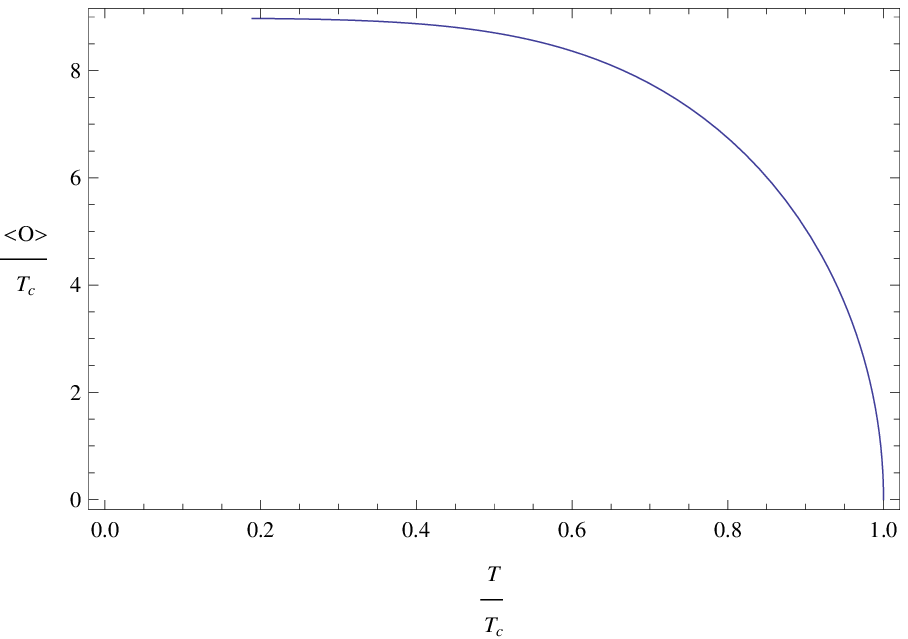}
\caption{The condensate of the 2D p-wave superconductors for the operator
$\mathcal{O}$ corresponding to D1/D5 (top-left), D2/D4 (top-right) and D3/D3
(bottom).}
\label{figure1}
\end{figure}

We plot in FIG.~\ref{figure1} the condensate $\langle\mathcal{O}\rangle$ versus dimensionless temperature $T/T_c$. Explicitly, they approach fixed constants as $T$ goes to zero, as is expected for a superconductor. However, the expectation values of the condensates near zero temperature are much larger than predictions from BCS theory, which should be explained as the strongly coupled feature of holographic superconductors. Moreover, our results are similar to higher dimensional counterparts \cite{0803.3295}. This may be understood as one of the universal properties of holographic methods applied to strongly interacting superconductors.

In the mean field theory for the superconductors, the order parameters have them square root behavior,
\begin{equation}
\langle\mathcal{O}\rangle\sim \left(T-T_c\right)^{1/2} \quad\text{when}\quad
T\rightarrow T_c.
\end{equation}
By fitting the curves in FIG.~\ref{figure1}, we also find such mean field behavior for the condensates in our models:
\begin{equation}
D1/D5:\quad \langle \mathcal{O}\rangle\approx
(17.8815T_c)^2\left(1-T/T_c\right)^{1/2} \quad\text{as}\quad T\rightarrow T_c
\end{equation}
where the critical temperature when expressed in terms of the charge density is
$T_c=0.0654409\rho$;
\begin{equation}
D2/D4:\quad \langle\mathcal{O}\rangle\approx
15.6435T_c\left(1-T/T_c\right)^{1/2} \quad \text{when}\quad T\rightarrow T_c
\end{equation}
with $T_c=0.0692836\rho$;
\begin{equation}
D3/D3:\quad \langle\mathcal{O}\rangle\approx
15.8992T_c\left(1-T/T_c\right)^{1/2} \quad\text{as}\quad T\rightarrow T_c
\end{equation}
where the critical temperature $T_c$ is expressed in terms of the chemical potential as $T_c=0.0827962\mu$.

From these numerical results, we conclude that all three models nearly give the same results when concerned with superconducting condensates.

\begin{figure}[h]
\includegraphics[scale=0.8]{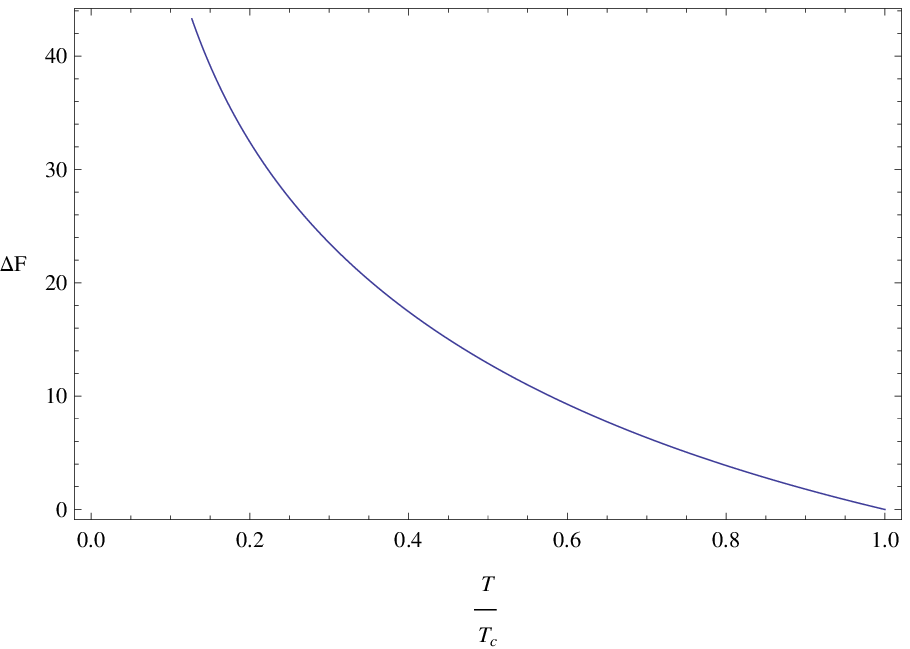}
\includegraphics[scale=0.8]{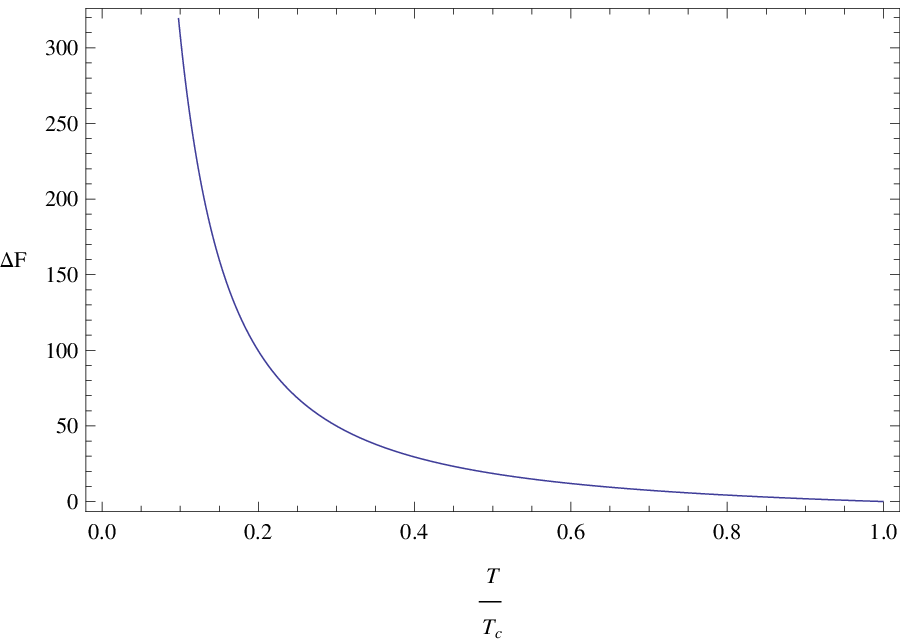}
\includegraphics[scale=0.8]{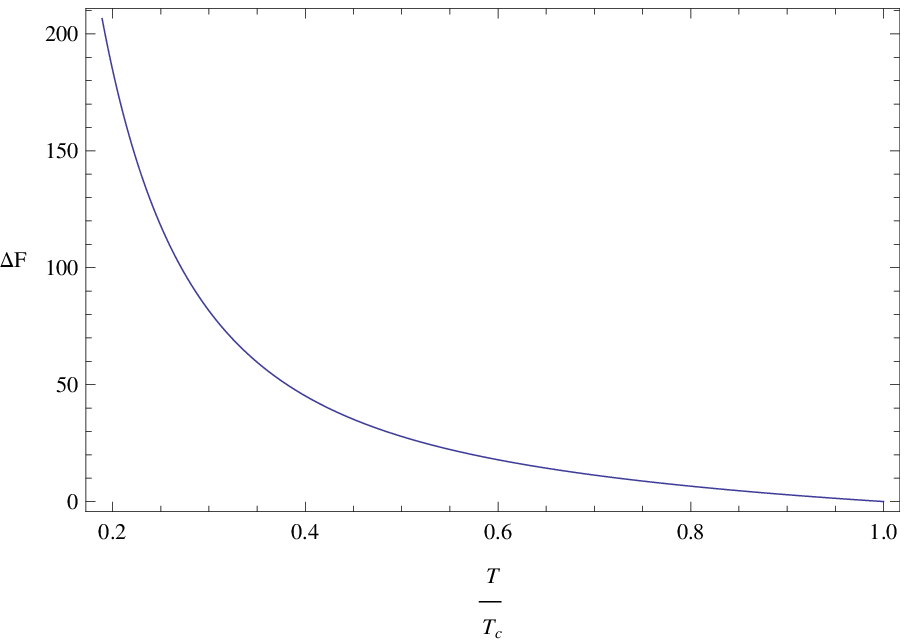}
\caption{The dimensionless free energy difference between the normal and the
superconducting phases for D1/D5 (top-left), D2/D4 (top-right) and D3/D3
(bottom), respectively.}
\label{figure2}
\end{figure}

Before closing this subsection, we plot the results for the free energies for the three systems, which can be taken as one evidence for that the superconducting phase transition does happen when decreasing the temperature to the critical value $T_c$. With the equations of motion (\ref{eq15},\ref{eq24},\ref{eq33}) for the backgrounds $\psi$ and $\phi$, we can reduce the action in eq.~(\ref{actionnew}) to some simpler expression. Then, the free energy difference between the normal and superconducting phases is
\begin{equation}
\Delta \Xi_{N-SC}=\mathcal{V}\int _0^1 du
\sqrt{-G}g^{xx}g^{tt}[\phi(u)\psi(u)]^2\equiv\mathcal{V} \Delta F,
\end{equation}
where $\mathcal{V}$ is a model dependent factor and we will not give its explicit expression here since it does not affect later arguments. FIG.~\ref{figure2} is for the plot of the free energy difference $\Delta F$.
We clearly see from these plots that the superconducting phases are thermodynamically favored below the critical temperature $T_c$.
\section{Electromagnetic fluctuation and AC conductivity}\label{section3}
\subsection{Fluctuation analysis: formula for the conductivity}
In this section we move on to the study of the electromagnetic response for the superconductor models constructed in previous section. In particular, we are concerned with the AC conductivity $\sigma(\omega)$ and assume that the fluctuations have no spatial dependence. We focus on the following decoupled sector $\left\{a_t^1(t,u),a_t^2(t,u),a_x^3(t,u)\right\}$ for the gauge field fluctuations and take the Fourier ansatz for them, say
$\left\{a_t^1(t,u),a_t^2(t,u),a_x^3(t,u)\right\}\sim e^{-i\omega
t}\left\{a_t^1(u),a_t^2(u),a_x^3(u)\right\}$. Then the linearized version of
eq.~(\ref{eq motion}) when considering the fluctuations are as follows,
\begin{equation} \label{eq:15}
\text{D1/D5}:\quad\left\{ \begin{aligned}
&{a_t^1}^{\prime\prime}- \frac{1}{u}{a_t^1}^{\prime}+\frac{u^2}{f(u)}\phi\psi
a_x^3=0\\
&{a_t^2}^{\prime\prime}- \frac{1}{u}{a_t^2}^{\prime}-\frac{u^2}{f(u)}
\left(i\tilde{\omega}\psi a_x^3+\psi^2 a_t^2\right)=0\\
&{a_x^3}^{\prime\prime}+\left[\frac{f^{\prime}(u)}{f(u)}-
\frac{1}{u}\right]{a_x^3}^{\prime}- \frac{u^2}{f^2(u)}\left[-\tilde{\omega}^2
a_x^3+\phi\psi a_t^1+ i \tilde{\omega}\psi a_t^2\right]=0,
\end{aligned} \right.
\end{equation}
where dimensionless frequency $\tilde{\omega}=\frac{3}{2\pi T}\omega$. For later convenience, we in the below list corresponding equations for D2/D4 and D3/D3 models,
\begin{equation} \label{eq:24}
\text{D2/D4}:\quad\left\{ \begin{aligned}
&{a_t^1}^{\prime\prime}+ \frac{u}{f(u)}\phi\psi a_x^3=0\\
&{a_t^2}^{\prime\prime}- \frac{u}{f(u)}\left(i\tilde{\omega}\psi a_x^3+\psi^2
a_t^2\right)=0\\
&{a_x^3}^{\prime\prime}+ \frac{f^{\prime}(u)}{f(u)}{a_x^3}^{\prime}-
\frac{u}{f^2(u)}\left[-\tilde{\omega}^2 a_x^3+\phi\psi a_t^1+ i
\tilde{\omega}\psi a_t^2\right]=0,
\end{aligned} \right.
\end{equation}
with $\tilde{\omega}=\frac{5}{4\pi T}\omega$;
\begin{equation} \label{eq:33}
\text{D3/D3}:\quad\left\{ \begin{aligned}
&{a_t^1}^{\prime\prime}+\frac{1}{u}{a_t^1}^{\prime}+\frac{1}{f(u)}\phi\psi
a_x^3=0\\
&{a_t^2}^{\prime\prime}+\frac{1}{u}{a_t^2}^{\prime}-\frac{1}{f(u)}
\left(i\tilde{\omega}\psi a_x^3+\psi^2 a_t^2\right)=0\\
&{a_x^3}^{\prime\prime}+
\left[\frac{f^{\prime}(u)}{f(u)}+\frac{1}{u}\right]{a_x^3}^{\prime}-\frac{1}{
f^2(u)}
\left[-\tilde{\omega}^2 a_x^3+\phi\psi a_t^1+ i \tilde{\omega}\psi
a_t^2\right]=0,
\end{aligned} \right.
\end{equation}
where $\tilde{\omega}=\frac{1}{\pi T}\omega$. Actually, these modes are not independent because the radial gauge $a_u^a=0$ has been chosen in deriving these fluctuation equations. This gauge choice gives two constraint equations for this decoupled sector. We do not present them here as these constraints have no effect on the definition as well as the numerical computations for the conductivity, which will be clear later. The AC conductivity is defined by following Kubo's formula,
\begin{equation}
\sigma(\omega)=\frac{G_{R}(\omega)}{i\omega}.
\end{equation}
Therefore, our aim to produce the electromagnetic response of the superconducting models we constructed in section \ref{section2} is then reduced to calculation of the retarded Green's function $G_R(\omega)$. Under gauge/gravity duality approach, a good prescription for the retarded Green's function can be found in \cite{hep-th/0205051}.

For the retarded Green's function, we should impose ingoing wave boundary condition at the horizon for these fluctuation modes. A simple Frobenius analysis for eqs.~(\ref{eq:15} ,\ref{eq:24},\ref{eq:33}) near the horizon reveals that
\begin{equation} \label{eq horizon 152433}
\left\{ \begin{aligned}
&a_x^3=(1-u)^{\alpha}\left[1+{a_x^3}^{(1)}(1-u)+{a_x^3}^{(2)}(1-u)^2+{a_x^3}^
{(3)}(1-u)^3+\cdots\right]\\
&a_t^1=(1-u)^{\alpha}\left[{a_t^1}^{(1)}(1-u)+{a_t^1}^{(2)}(1-u)^2+{a_t^1}^
{(3)}(1-u)^3+\cdots\right]\\
&a_t^2=(1-u)^{\alpha}\left[{a_t^2}^{(1)}(1-u)+{a_t^2}^{(2)}(1-u)^2+{a_t^2}^
{(3)}(1-u)^3+\cdots\right],\end{aligned} \right.
\end{equation}
where we have used the linearity of eqs.~(\ref{eq:15},\ref{eq:24},\ref{eq:33}) to set the scale of $a_x^3$ to 1 at the horizon. The indices appearing in these equations are $-i\tilde{\omega}/6$, $-i\tilde{\omega}/5$ and $-i\tilde{\omega}/4$ for the three models, respectively. The coefficients in above equations can be uniquely determined once plugging these expansions into eqs.~(\ref{eq:15},\ref{eq:24},\ref{eq:33}) and counting by powers of $(1-u)$. Therefore, eqs.~(\ref{eq horizon 152433}) can provide initial conditions for these second order differential equations. We in fact use these power solutions to do numerical integration from the horizon to the conformal boundary by the
mathematica NDSolve.

At the AdS conformal boundary $u=0$, the general solution to the equations of motion is distinguished between the D1/D5 (D2/D4) and D3/D3 models. More specifically,
\begin{equation}\label{eq bdy 1524}
\text{D1/D5 (D2/D4)}:\quad\left\{ \begin{aligned}
&a_t^1={A_t^1}^{(0)}+ {A_t^1}^{(1)} u^2 (u)+\cdots\\
&a_t^2={A_t^2}^{(0)}+ {A_t^2}^{(1)} u^2 (u)+\cdots\\
&a_x^3={A_x^3}^{(0)}+ {A_x^3}^{(1)} u^2 (u)+\cdots
\end{aligned}\right.
\end{equation}
and
\begin{equation}\label{eq bdy 33}
\text{D3/D3}:\quad\left\{ \begin{aligned}
&a_t^1={A_t^1}^{(0)}+ {A_t^1}^{(1)} \log u+\cdots\\
&a_t^2={A_t^2}^{(0)}+ {A_t^2}^{(1)} \log u+\cdots\\
&a_x^3={A_x^3}^{(0)}+ {A_x^3}^{(1)} \log u+\cdots
\end{aligned}\right.
\end{equation}
where we have represented the boundary expansions for the D1/D5 and D2/D4 models together and the expressions in the parenthesis are for the D2/D4 system. Explicitly, the logarithmic terms appear again in the boundary behavior of the gauge field fluctuations for the D3/D3 model.

As argued in \cite{0805.2960}, the conductivity is a physical quantity and should be gauge invariant. We should construct a new mode from $a_t^1, a_t^2, a_x^3$ and this mode should be invariant under the gauge transformation that keeps our gauge choice. The details for the constructions of the gauge invariant modes can be found in the original paper \cite{0805.2960} and we in the following write down this mode directly,
\begin{equation}
\tilde{a_x^3}\equiv a_x^3+\psi\frac{i \tilde{\omega} a_t^2+\phi
a_t^1}{\phi^2-\tilde{\omega}^2}.\label{new mode}
\end{equation}
Plugging the boundary behavior in eq.~(\ref{eq bdy 1524}) into the newly defined mode in eq.~(\ref{new mode}) and expanding it near $u=0$ results in
\begin{equation}
\text{D1/D5(D2/D4)}: \quad \tilde{a_x^3}=\tilde{A_x^3}^{(0)}+
\tilde{A_x^3}^{(1)}u^2 (u)+\cdots,
\end{equation}
with
\begin{equation}
\tilde{A_x^3}^{(0)}={A_x^3}^{(0)},\quad
\tilde{A_x^3}^{(1)}={A_x^3}^{(1)}+\psi^{(1)}\frac{i \tilde{\omega}{A_t^2}^{(0)}
+\mu{{A_t^1}^{(0)}}}{\mu^2-\tilde{\omega}^2}.\label{new bdy}
\end{equation}
Then, the formula for the conductivity can be straightforwardly written down for D1/D5 and D2/D4 systems,
\begin{equation}
\sigma(\omega)=\frac{1}{i\omega}\frac{\tilde{A_x^3}^{(1)}}{\tilde{A_x^3}^{(0)}}.
\label{conductivity}
\end{equation}
Notice that the above formula is the same as the equation (4.19) of \cite{0805.2960}. We can expect that numerical results for the conductivity of these two models should have some similarities with the results reported in \cite{0805.2960} for $\tilde{\sigma}_{xx}$ and this does happen for our results.

Involving the D3/D3 model, we have the corresponding results for the mode $\tilde{a_x^3}$ near $u=0$,
\begin{equation}
\text{D3/D3}: \quad \tilde{a_x^3}=\tilde{A_x^3}^{(1)}+ \tilde{A_x^3}^{(0)}\log u
+\cdots,
\end{equation}
with
\begin{equation}
\tilde{A_x^3}^{(0)}={A_x^3}^{(0)},\quad
\tilde{A_x^3}^{(1)}={A_x^3}^{(1)}+\frac{{A_t^1}^{(0)}\psi^{(1)}}{\mu}.
\end{equation}
Due to the appearance of the logarithmic term $\log u$ in the asymptotic behavior for the fluctuation modes, we should identify $\tilde{A_x^3}^{(0)}$ as the source and $\tilde{A_x^3}^{(1)}$ as the expectation value of the dual operator. Moreover, the definition for the retarded Green's function as well as the conductivity should be modified to
\begin{equation}
G_R(\omega)=-\frac{\tilde{A_x^3}^{(1)}}{\tilde{A_x^3}^{(0)}}\quad
\text{and}\quad
\sigma(\omega)=-\frac{1}{i\omega}\frac{\tilde{A_x^3}^{(1)}}{\tilde{A_x^3}^{(0)}}
.
\end{equation}
We have seen that this formula is greatly different from the counterparts
eqs.~(\ref{new bdy},\ref{conductivity}) for D1/D5 and D2/D4 models. In the next subsection, we will see that this difference will be reflected in the imaginary parts of the conductivity.
\subsection{Numerical results for the conductivity}
We report our numeric results for the AC conductivity for all three models in this subsection. Before this, we have a brief explanation for our numerical method. We use the horizon expansions as in eq.~(\ref{eq horizon 152433}) to generate initial conditions for the equations of motion for the fluctuations. We then use the mathematica NDSolve to numerically solve these eqs.~(\ref{eq:15},\ref{eq:24},\ref{eq:33}). The expansion coefficients appearing in the conductivity formula can be directly read off from boundary behavior of different modes as in eqs.~(\ref{eq bdy 1524},\ref{eq bdy 33}) once the
backgrounds $\psi$ and $\phi$ as well as the frequency $\omega$ are specified.

\begin{figure}[h]
\includegraphics[scale=0.8]{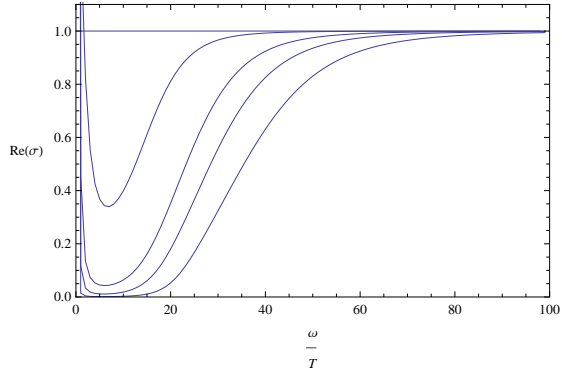}
\caption{The real part of the AC conductivity with different condensates
corresponding to $T/T_c=1.0, 0.603055, 0.277291, 0.193113, 0.126896$ for D1/D5
from top to down.}
\label{figure3a}
\end{figure}

\begin{figure}[h]
\includegraphics[scale=0.8]{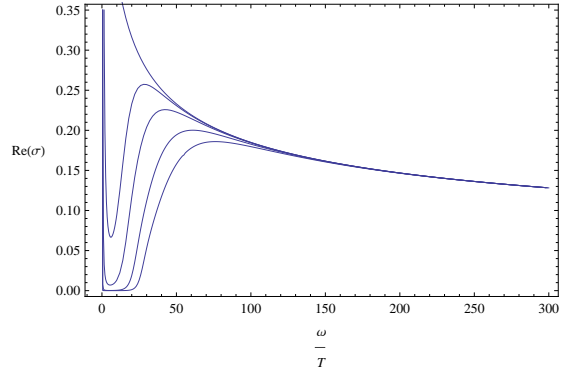}
\caption{The real part of the AC conductivity with different condensates
corresponding to $T/T_c=1.0, 0.464611, 0.254859, 0.141036, 0.0978091$ for D2/D4
model from top to down.}
\label{figure3b}
\end{figure}

\begin{figure}[h]
\includegraphics[scale=0.8]{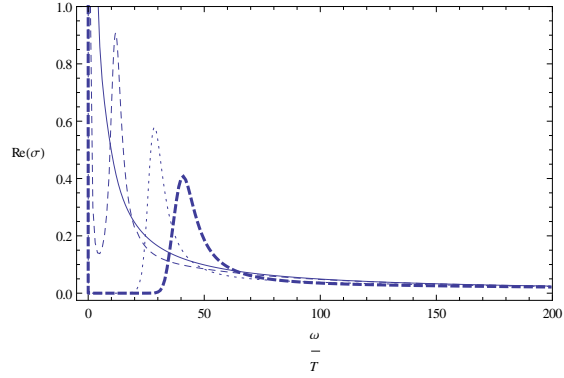}
\caption{The real part of the AC conductivity with different condensates
corresponding to $T/T_c=1.0(\text{Solid}), 0.664919(\text{Dashed}),
0.269669(\text{Dotted}), 0.189074(\text{Thick Dashed})$ for D3/D3 model.}
\label{figure3c}
\end{figure}

\begin{figure}[h]
\includegraphics[scale=0.8]{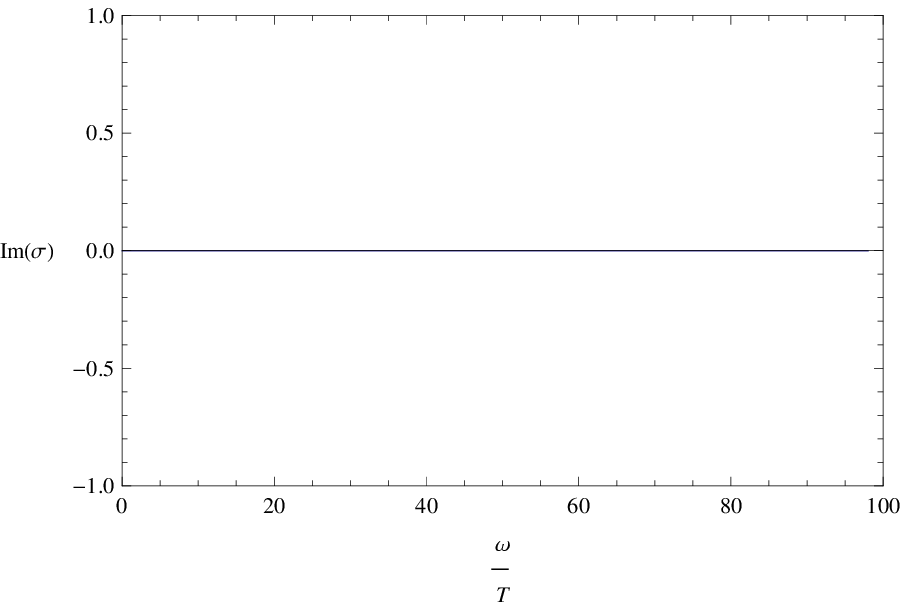}
\includegraphics[scale=0.8]{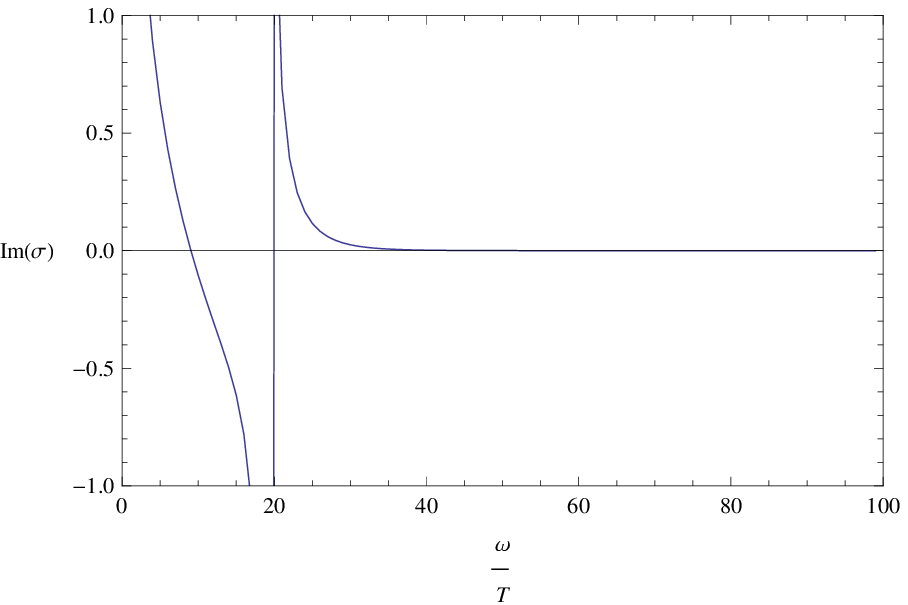}
\includegraphics[scale=0.8]{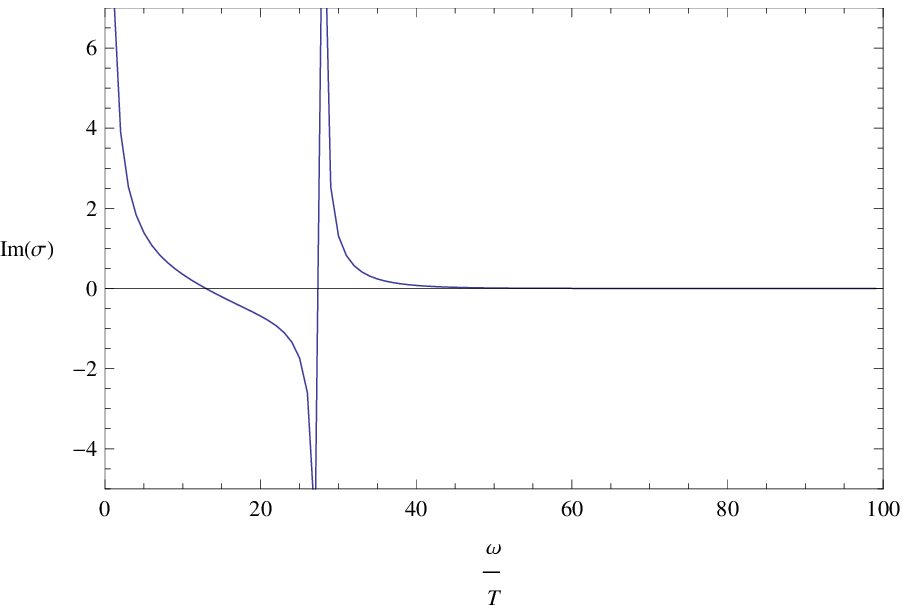}
\includegraphics[scale=0.8]{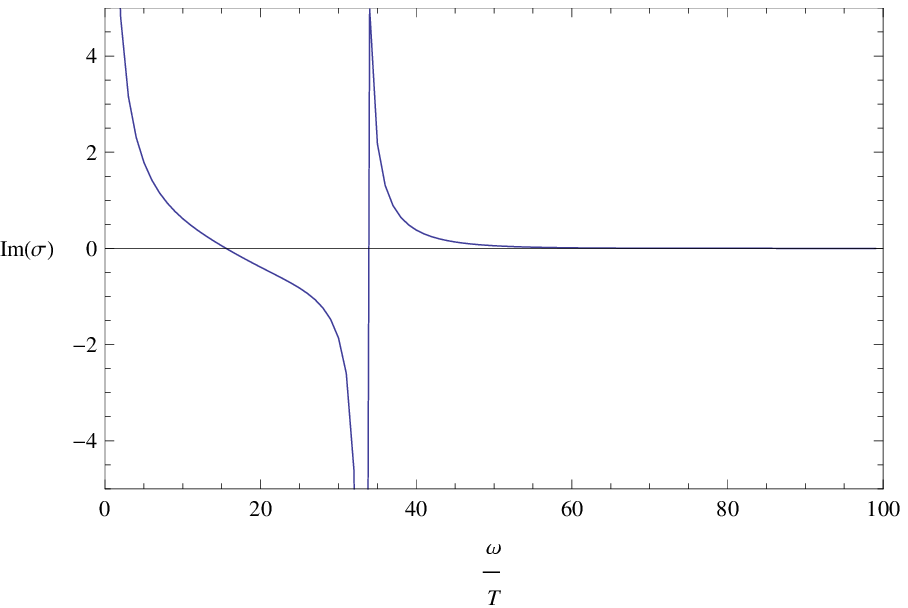}
\caption{The imaginary part of the AC conductivity with different condensates
corresponding to $T/T_c=1.0, 0.603055, 0.277291, 0.193113, 0.126896$ for D1/D5
from left to right and top to down.}
\label{figure4}
\end{figure}

\begin{figure}[h]
\includegraphics[scale=0.8]{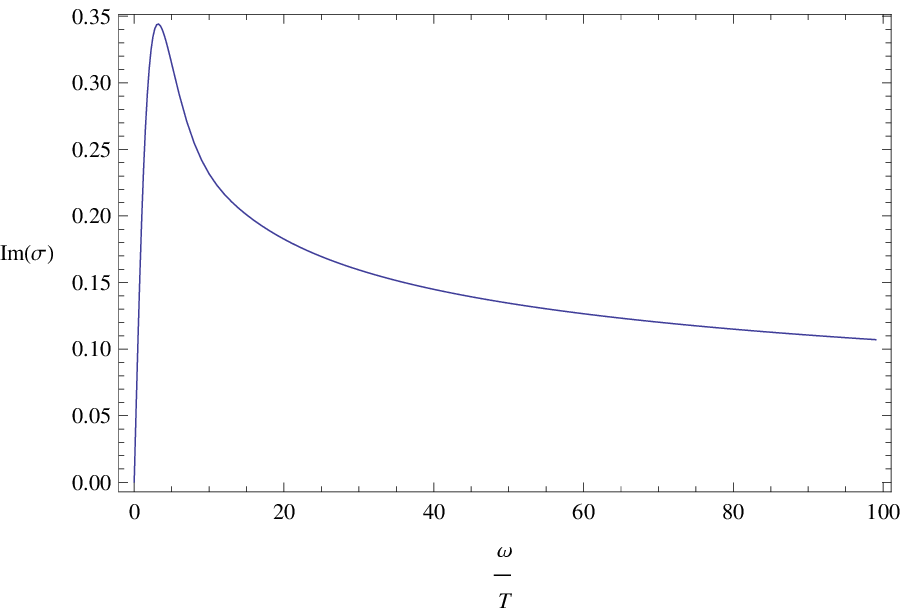}
\includegraphics[scale=0.8]{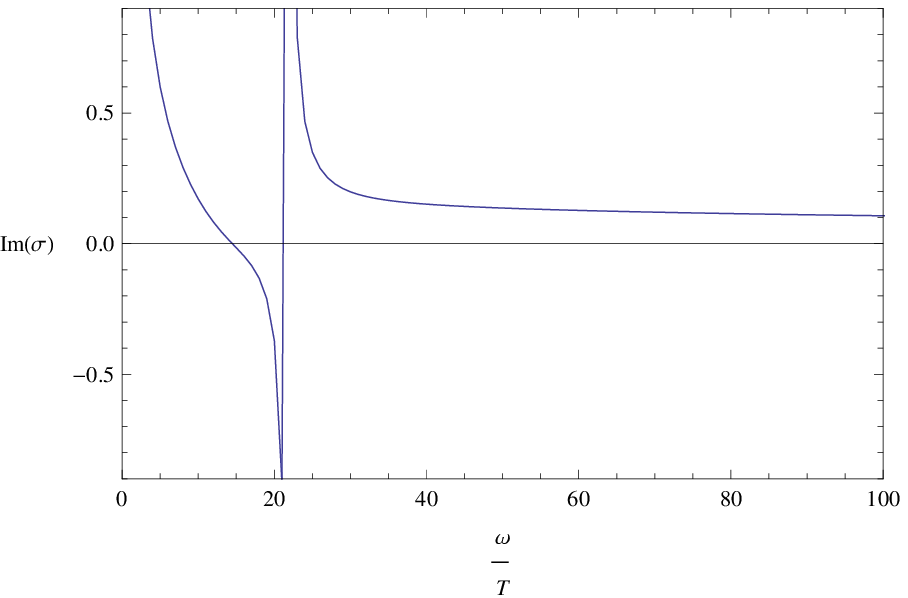}
\includegraphics[scale=0.8]{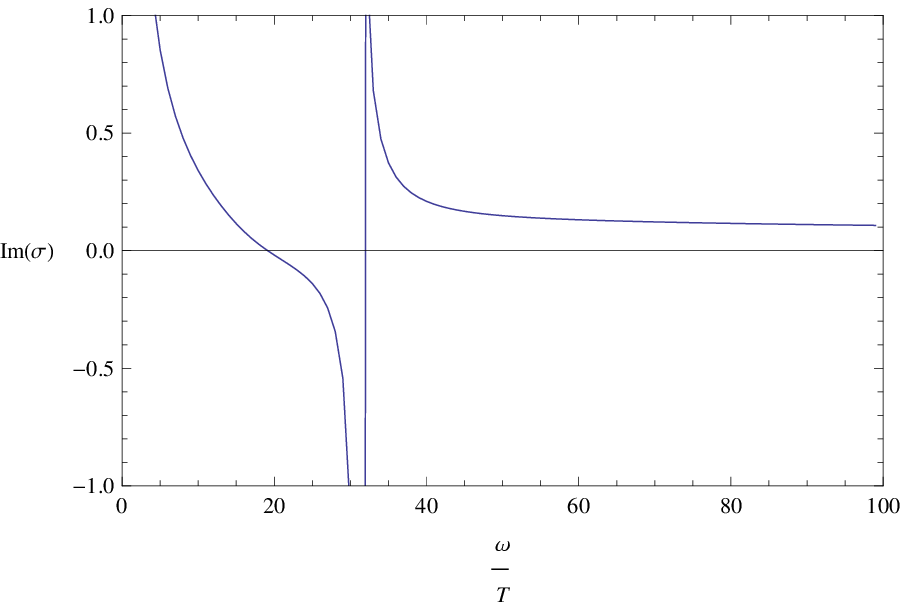}
\includegraphics[scale=0.8]{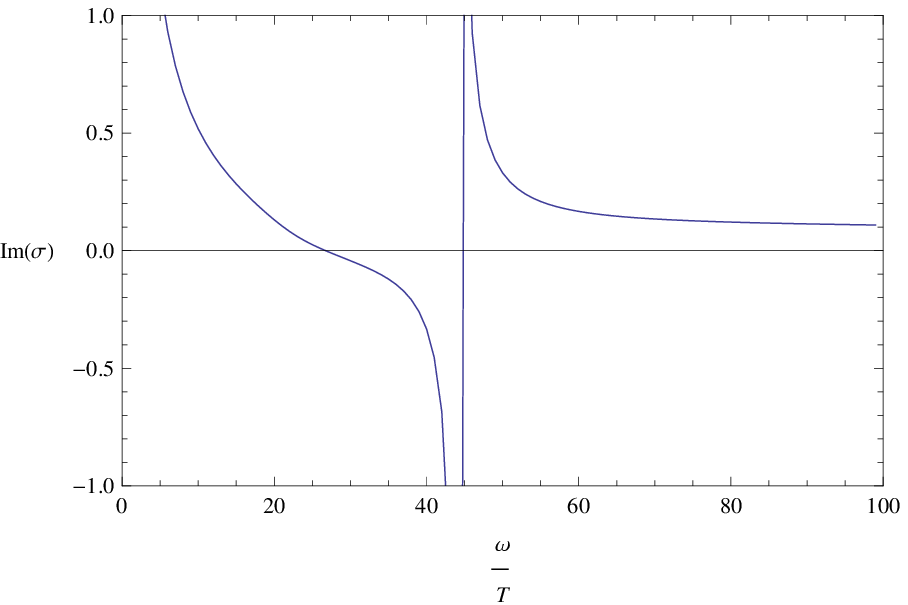}
\caption{The imaginary part of the AC conductivity with different condensates
corresponding to $T/T_c=1.0, 0.464611, 0.254859, 0.141036$ for D2/D4 from left
to right and top to down.}
\label{figure5}
\end{figure}

\begin{figure}[h]
\includegraphics[scale=0.8]{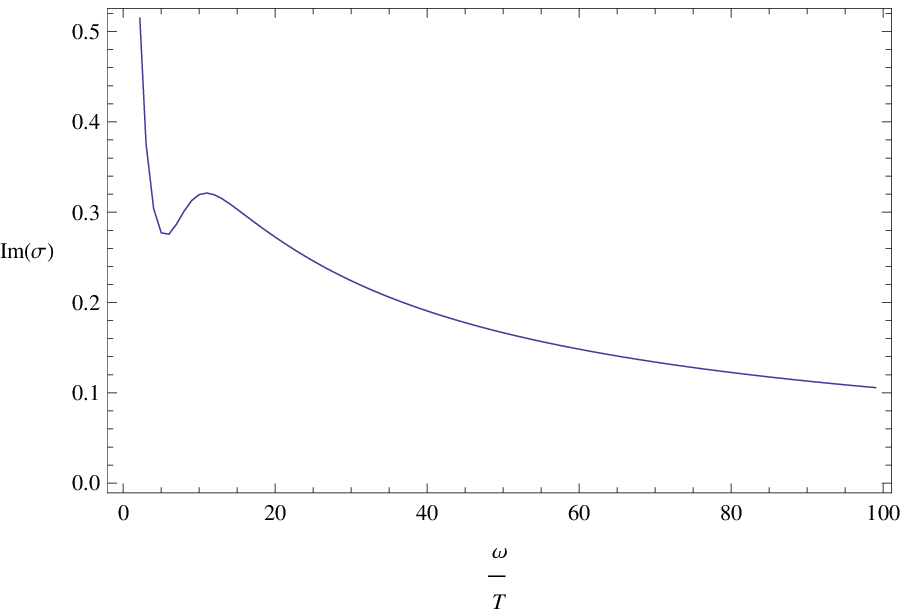}
\includegraphics[scale=0.8]{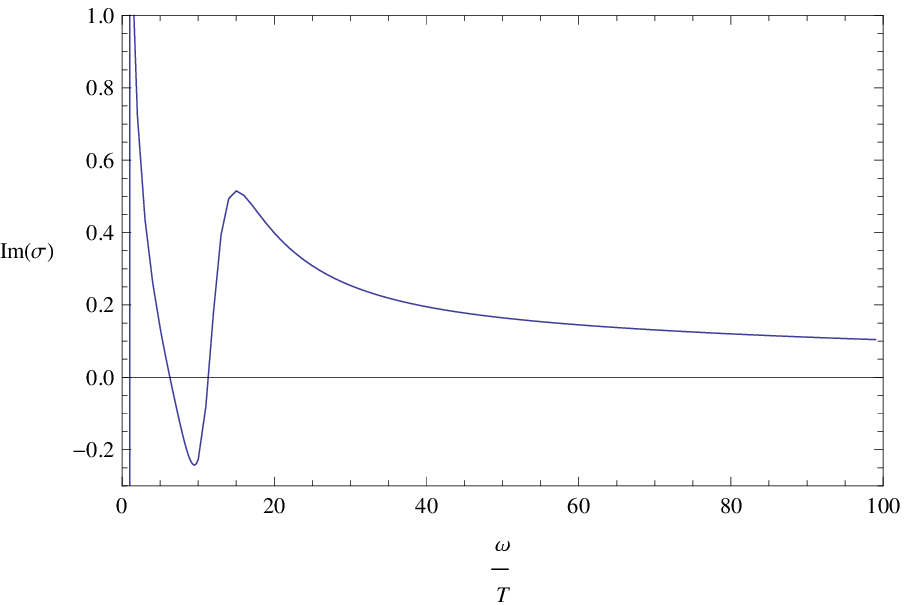}
\includegraphics[scale=0.8]{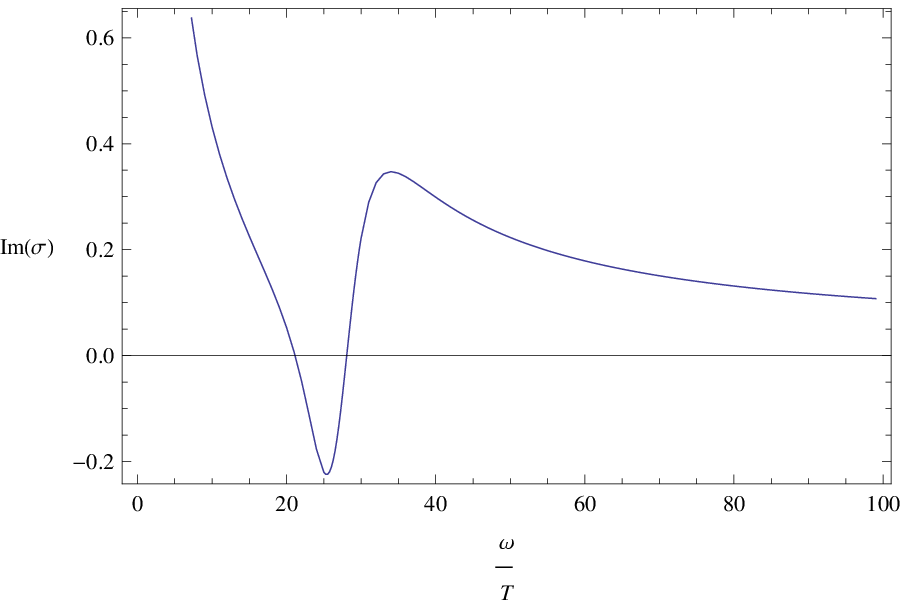}
\includegraphics[scale=0.8]{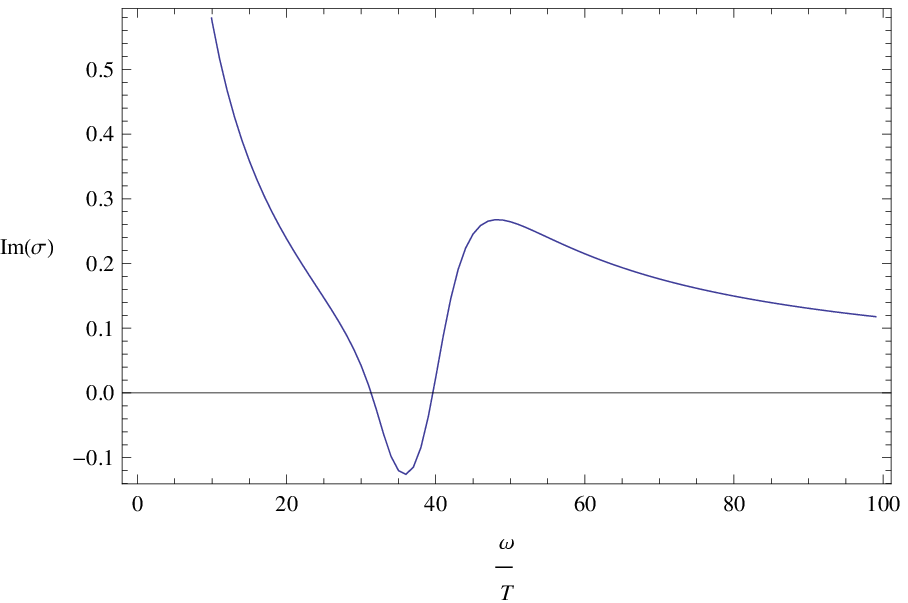}
\caption{The imaginary part of the AC conductivity with different condensates
corresponding to $T/T_c=1.0, 0.664919, 0.269669, 0.189074$ for D3/D3 from left
to right and top to down.}
\label{figure6}
\end{figure}

In FIG.~(\ref{figure3a},\ref{figure3b},\ref{figure3c},\ref{figure4},\ref{figure5},
\ref{figure6}), we plot the real and imaginary parts of the AC conductivities for all the three models.

We conclude with some explanations of these figures for the conductivity:

Firstly consider the normal phase, i.e. $T/T_c=1$. For the D1/D5 model, $\text{Re}\sigma(\omega)$ is a constant 1 and $\text{Im}\sigma(\omega)$ is always zero. This is consistent with higher dimensional models. On the other hand, for D2/D4 and D3/D3 models, $\text{Re}\sigma(\omega)$ decreases monotonously as increasing the frequency $\omega/T$. Moreover, $\text{Re}\sigma(\omega)$ of D3/D3 approaches to zero at high frequency while $\text{Re}\sigma(\omega)$ for D2/D4 goes to a nonzero constant. This feature has also been found by studying the flavor U(1) dynamics of D3/D3 model or BTZ black hole in \cite{0909.3526,0909.4051,1008.3904}.

When decreasing the temperature, we go to the condensed phase, i.e.
$\langle\mathcal{O}\rangle\neq0$. We find that $\text{Re}\sigma(\omega)$ for all three models display gap formation, which is similar to the findings in \cite{0803.3295,*0810.1563,0805.2960,1008.3904}. More specifically, $\text{Re}\sigma(\omega)$ is very small in the infrared (i.e. for small $\omega/T$ in the figures) and then rises quickly at some critical frequency $\omega_g$. Near $\omega=0$, there appears a delta peak in the conductivity, which can be explained by the Kramers-Kronig relations. This relation relates the real and imaginary parts of casual quantities and can detect the distributional parts of them. Take the D1/D5 system as an example. The Kramers-Kronig relation states that,
\begin{equation}
\text{Im}[\sigma(\omega)]=-P\int_{-\infty}^{\infty}\frac{d\omega^{\prime}}{
\omega}
\frac{\text{Re}[\sigma(\omega^{\prime})]}{\omega^{\prime}-\omega}
\label{kkequation}
\end{equation}
where P denotes the principal part of the integration. It is clear from this formula that $\text{Re}(\omega)$ has a delta peak, $\text{Re}\sigma(\omega)\sim \delta(\omega)$, only when $\text{Im}\sigma(\omega)$ has a pole at $\omega=0$, $\text{Re}\sigma(\omega)\sim 1/\omega$, and vice versa. The delta peak at $\omega=0$ indicates a DC superconductivity. Similarly, the pole in the imaginary part of the AC conductivity at finite $\omega$ can also be understood from the Kramers-Kronig relation. As mentioned before, this pole can be easily seen from the formula for the conductivity eq.~(\ref{new bdy}). From eq.~(\ref{kkequation}), we can conclude that a simple pole in Im[$\sigma(\omega)$] at
$\omega=\omega_0$ implies a delta-function contribution $\delta(\omega-\omega_0)$ to Re[$\sigma(\omega)$]. We already see that this peak becomes more higher and narrower as decreasing the temperature. However, as for as we understand, we do not think of it as a massive excitation. We rather take it as artificial as we use the newly defined mode to plot the conductivity, which in some sense implies that it is not a good choice of gauge invariant mode used in this work and the pole is introduced by hand, however inevitable as far as we know.

In the superconducting phase, there is a pole in $\text{Im}\sigma(\omega)$ for D1/D5 and D2/D4 models, which is in contrast to the D3/D3 model. Actually, from the formulae of the conductivity for all three models, we have seen some differences between them and the peak should appear at $\tilde{\omega}_0=\mu$ from eq.~(\ref{new bdy}) with the residue proportional to the condensate $\langle\mathcal{O}\rangle$. However, for D3/D3 model, we can easily read off the $\omega_g$ if we follow the analysis of \cite{0810.1077} and define $\omega_g$ as the frequency which minimizes $\text{Im}(\omega)$,
\begin{equation}
\frac{\omega_g}{T}\approx 25 \quad \text{when} \quad \frac{T}{T_c}=0.269669.
\end{equation}

When taking large $\omega$ limit, all the results go to those of the normal phase, which is expected from general grounds, the large $\omega$ will wash out the effect of the superconducting condensates. This can also be seen clearly from the eqs.~(\ref{eq:15},\ref{eq:24},\ref{eq:33}).

With these comments in mind, we can conclude that all three models can reproduce some basic features of the 1+1 dimensional p-wave superconductor, like the DC infinite conductivity (a signal for superconductivity) and gap formation when decreasing the temperature. The results of D1/D5 model are more related to the higher dimensional counterparts. The D3/D3 model for p-wave superconductor have some common feature with its s-wave case as studied in \cite{1008.3904}.

\section{Summary}\label{section4}
In this paper we have taken the D-brane probe approach to explore some properties of holographic p-wave superconductor in 1+1 dimensional spacetime. In the large N limit, we can bypass the no-go theorem (the CMW theorem) which forbids the superconducting phase transition. We found that all the three models are quantitatively similar in producing some key features of the p-wave superconductor, like the mean field behavior of the superconducting condensate near $T_c$, the DC delta peak and the gap formation, etc. These are common with the higher dimensional superconductors under the holographic approach. Therefore, these characteristics can be regarded as universality of the holographic method. Besides these, we also found some particular ones for different models, especially for the AC conductivity. Specially, the high frequency limits of the conductivities are different: for D1/D5, it goes like the higher dimensional counterparts, $\text{Re}\sigma(\omega)\rightarrow 1$ and $\text{Im}\sigma(\omega)\rightarrow 0$; for D2/D4, both $\text{Re}\sigma(\omega)$ and $\text{Im}\sigma(\omega)$ approaches some nonzero constants but very slowly compared to the other two models; the conductivity of the D3/D3 model is basically the same as its s-wave counterpart, which is first studied in \cite{1008.3904}. However, the condensates found here is greatly different from the results of \cite{1008.3904} where $\langle\mathcal{O}\rangle$ decreases or increases in the zero temperature limit. Leaving aside these model dependent differences, we can conclude the holographic approach to the
superconducting phase transition can reproduce some common features and give us some directions for strongly correlated condensed matter system.

Actually, we in this work only explored some basic aspects of the 1+1-dimensional superconductor from holographic viewpoint. There are some interesting problems deserved further investigations. The first one is to go beyond the zero quark mass limit and study its effect on the superconducting phase transition and conductivity. There is only one conductivity for 1+1-dimensional spacetime and it looks like $\sigma_{xx}(\omega)$ in the higher dimensional case. However, such quantity in colorful superconductors from probe
D-brane approach has not been computed. So we hope to come to this question in the near future and have a complete comparison between our work and its higher dimensional counterparts.
\begin{acknowledgments}
The author greatly thanks Johanna Erdmenger, Jonathan Shock, Rene Meyer, Xin Gao and Qiaoni Chen for useful discussions. This work was supported by the Joint MPS-CAS Doctoral Training program.
\end{acknowledgments}

\end{document}